\pdfoutput=1
\makeatletter
\@ifundefined{@parse@version@dash}{%
\def\@parse@version#1{\@parse@version@0#1}
\def\@parse@version@#1/#2/#3#4#5\@nil{%
\@parse@version@dash#1-#2-#3#4\@nil}
\def\@parse@version@dash#1-#2-#3#4#5\@nil{%
  \if\relax#2\relax\else#1\fi#2#3#4 }
}{}
\makeatother
\documentclass[aps,prl,reprint]{revtex4-2}

\usepackage{amsmath}          
\usepackage{mathrsfs}         
\usepackage{eqparbox}         
\usepackage{bbold}            
\usepackage{physics}          
\usepackage{graphicx}         
\usepackage{verbatim}         
\usepackage{color}            
\usepackage{subfigure}        
\usepackage{hyperref}         
\usepackage{gnuplot-lua-tikz} 

\usepackage{tikz}
\usetikzlibrary{arrows,shapes}
\usetikzlibrary{matrix,arrows}
\usetikzlibrary{positioning}
\usetikzlibrary{calc,through}
\usetikzlibrary{decorations.pathmorphing}
\usetikzlibrary{decorations.markings}

\tikzset{
  fermion/.style={postaction={decorate},
    decoration={markings,mark=at position .55 with {\arrow{>}}}},
  fermionbar/.style={postaction={decorate},
    decoration={markings,mark=at position .55 with {\arrow{<}}}},
}

\newcommand{\eqmathbox}[2][T]{\eqmakebox[#1][l]{$#2$}}

\newcommand{\sigmabar}{{\overline{\sigma}}}

\newcommand{\lagrangian}[1]{\mathscr{L}_{\mathrm{#1}}}

\newcommand{\leftFermi}[2]{{#1}^\dagger_{L} (\overline{\sigma} \cdot {#2}) {#1}^{\phantom{\dagger}}_{L}}
\newcommand{\rightFermi}[2]{{#1}^\dagger_{R} (\sigma \cdot {#2}) {#1}^{\phantom{\dagger}}_{R}}

\newcommand{\leftMass}[2]{{#1}^\dagger_{L} {#2}^{\phantom{\dagger}}_{R}}
\newcommand{\rightMass}[2]{{#1}^\dagger_{R} {#2}^{\phantom{\dagger}}_{L}}

\newcommand{\fermiQuarticLeft}[2]{{#1}^\dagger_L {#2}^{\phantom{\dagger}}_R {#2}^\dagger_R {#1}^{\phantom{\dagger}}_L}
\newcommand{\fermiQuarticLeftVar}[2]{{#1}^\dagger_L {#1}^{\phantom{\dagger}}_R {#2}^\dagger_L {#2}^{\phantom{\dagger}}_R}
\newcommand{\fermiQuarticLeftVarPrime}[2]{{#1}^\dagger_L {#2}^{\phantom{\dagger}}_R {#2}^\dagger_L {#1}^{\phantom{\dagger}}_R}
\newcommand{\fermiQuarticRight}[2]{{#1}^\dagger_R {#2}^{\phantom{\dagger}}_L {#2}^\dagger_L {#1}^{\phantom{\dagger}}_R}
\newcommand{\fermiQuarticRightVar}[2]{{#1}^\dagger_R {#1}^{\phantom{\dagger}}_L {#2}^\dagger_R {#2}^{\phantom{\dagger}}_L}
\newcommand{\fermiQuarticRightVarPrime}[2]{{#1}^\dagger_R {#2}^{\phantom{\dagger}}_L {#2}^\dagger_R {#1}^{\phantom{\dagger}}_L}

\newcommand{\phiZero}{\phi^{\phantom{\dagger}}_{0}}
\newcommand{\phiZeroDagger}{\phi^{\dagger}_{0}}
\newcommand{\phiPlus}{\phi^{\phantom{\dagger}}_{+}}
\newcommand{\phiMinus}{\phi^{\phantom{\dagger}}_{-}}

\newcommand{\shortMomentumMeasure}[1]{\widetilde{d{#1}}}

\newcommand{\shortDelta}[1]{\tilde{\delta}(#1)}

\DeclareMathOperator{\diag}{diag}

\newcommand{\fderiv}[2]{\frac{\delta {#1}}{\delta {#2}}}
\newcommand{\fderivTwo}[3]{\frac{\delta^2 {#1}}{\delta {#2} \delta{#3}}}

\newcommand{\fderivText}[2]{{\delta {#1}}/{\delta {#2}}}

\begin{document}

\title{Excitations of the Higgs boson}
\author{Gregory A. Wright}
\email[email: ]{Gregory.Wright@nokia-bell-labs.com}
\affiliation{Nokia Bell Laboratories, %
  600 Mountain Avenue, Murray Hill, NJ 07974}

\date{July 12, 2024}

\begin{abstract}
In a simplified version of the standard model with a single quark
doublet, I derive a transcendental equation for the complex Higgs
boson mass. The equation involves a divergent integral which is
regularized and renormalized conventionally.  Setting the Higgs
mass to its observed value, the decay width is narrow and
within experimental limits. The mass equation has additional roots on
other sheets in the complex energy plane. The first two are at 186
and 219 GeV. The lightest excitations are well defined though
wider than the 125 GeV Higgs because decays to $W^{+}W^{-}$ and $ZZ$
are allowed.
\end{abstract}

\maketitle

In this letter I describe a simplified version of the standard model
in which electroweak symmetry is dynamically broken by a
gravitationally bound condensate of fermions.  The collective modes of
this condensate are three phase excitations, corresponding to
Goldstone bosons, and a massive amplitude excitation which is the
Higgs boson. The Goldstone bosons are absorbed by the usual Higgs
mechanism~\cite{higgs:1964}. In contrast to the standard model, where
the Higgs mass is a free parameter, here it is determined by a
transcendental equation with a well isolated smallest root and an
infinite number of additional solutions accumulating at $\sqrt{2} v$,
where $v$ is the Higgs vacuum expectation value.

This work extends the old idea that fermion mass generation occurs
analogously as in superconductivity~\cite{nambu:1961a,vaks:1961}.  The
new idea here is that initially massless fermions are bound in a
condensate by their mutual gravitational attraction.  Because the
gravitational force is both long-range and unscreened, even though it
is weak, it can cause the formation of a correlated ground state (as
long as cosmological considerations can be neglected).  Also, every
fermion is attracted to the condensate with a strength equal to its
radiatively corrected gravitational coupling.  This provides a natural
explanation for why all fermions have masses.  Radiative corrections
allow the fermions to have different masses, even though the model is
invariant under $SU{(2)}_W \cross U{(1)}_Y$.  The model requires no
nonlinear gravitational effects, beyond the radiative corrections of
the graviton-fermion vertex by vector gauge loops, to produce the
effects described here, and in particular no third-order or higher
gravitational interaction.  A shortcoming of the model is that as yet
it provides no explanation for fermion flavor.

The model is defined by a lagrangian for two massless fermions, $t$
and $b$. In what follows, gauge fields play no role and are omitted.
I will use the momentum representation from the outset, with the
fermion fields in the Weyl representation. The integral of the
lagrangian is assumed to be the action within a functional integral.
The trace minus two Minkowski metric is used throughout and $\hbar = c =
1$.  The right and left-handed Weyl matrices are $\sigma^{\mu} = (\mathbb{1},
\boldsymbol\sigma)$, $\sigmabar^{\mu} = (\mathbb{1}, -\boldsymbol\sigma)$,
respectively, where $\mathbb{1}$ is the $2 \times 2$ identity matrix and
$\boldsymbol\sigma$ are the Pauli matrices. For brevity, the momentum
measure is written $\shortMomentumMeasure{p} = d^4p/{(2\pi)}^4$ and the
delta function in momentum space is $\shortDelta{p} = {(2\pi)}^4
\delta^{(4)}(p)$.

The kinetic part of the lagrangian is
\begin{equation}
  \label{eq:fermi-kinetc}
  \begin{aligned}
    \lagrangian{fermi} = \, & \leftFermi{t}{p} + \rightFermi{t}{p} + \\
                            & \leftFermi{b}{p} + \rightFermi{b}{p} \, ,
  \end{aligned}
\end{equation}
where from now on the two fermions will be referred to as the top and
bottom quarks. The validity of this interpretation will be shown.

As in the Bardeen-Cooper-Schrieffer (BCS) model of
superconductivity~\cite{bcs:1957}, an effective quartic
interaction approximates the interaction between the
fermions:
\begin{equation}
  \label{eq:quartic-lagrangian}
  \begin{split}
    \lagrangian{\mathrm{quartic}} = \, -f_t^2 \Bigl[
    \alpha \, \bigl( & \fermiQuarticRight{t}{t} +
                  \fermiQuarticLeft{b}{t} \bigr)+ \\
    \beta \, \bigl( & \fermiQuarticLeftVar{b}{t} +
                  \fermiQuarticRightVar{t}{b} \\
            -\, & \fermiQuarticRightVarPrime{b}{t}  -
                  \fermiQuarticLeftVarPrime{b}{t} \bigr) + \\
    \gamma \, \bigl( & \fermiQuarticLeft{b}{b} +
                  \fermiQuarticRight{b}{t} \bigr) \Bigr] \, .
  \end{split}
\end{equation}
The three terms in parentheses are separately invariant under
$SU{(2)}_W \cross U{(1)}_Y$.  Because the action is a lorentz scalar,
pairs of fermi fields must have opposite handedness (e.g., $t^\dagger_L
b_R$).  For the moment, the parameters $\alpha$, $\beta$ and $\gamma$ are arbitrary,
dimensionless numbers.  The overall coupling constant $f_t^2$ is the
product of a dimensionless ratio of momenta (two factors of momentum
in the numerator from the quark energy-momentum tensors and a momentum
squared in the denominator from the graviton propagator) and the
gravitational constant with dimension $-2$.  Following BCS, the
momentum dependence of the coupling will be ignored: $f_t^2$ is taken
as constant up to some cutoff, above which it is zero.  Taking the
coupling to be momentum independent is equivalent to $s$-wave pairing
of the condensate. Each pair of Weyl spinors has color indices
contracted to form a color singlet.  Color indices are suppressed
throughout but the color factor is provided in the final result.

\begin{figure}[!h]
  \begin{minipage}[c]{.45\linewidth}
    \begin{tikzpicture}[scale=0.4]
      \begin{scope}
        \clip (1, 0) rectangle (10, -1) ;
        \draw[thick] (1,0) -- (10,0) ;
        \foreach \x in {0, 0.25, ..., 11}
        \draw (\x, -1) -- ($ (\x, 0) + (1, 0) $) ;
      \end{scope}
      \draw[thick, decorate, decoration=snake, double] (4.5, 0) -- (4.5, 3) ;
      \draw[thick, decorate, decoration=snake, double] (6.5, 0) -- (6.5, 3) ;
      \draw[fermionbar, very thick] (2, 2) node [anchor=east] {$t^\dagger_R$} -- (4.5, 3) ;
      \draw[fermion,    very thick] (2, 4) node [anchor=east] {$t_R$}   -- (4.5, 3) ;
      \draw[fermion,    very thick] (9, 2) node [anchor=west] {$b_L$}   -- (6.5, 3) ;
      \draw[fermionbar, very thick] (9, 4) node [anchor=west] {$b^\dagger_L$} -- (6.5, 3) ;

      \filldraw[fill=black!50] (4.5, 3) circle (8pt) ;
      \filldraw[fill=black!50] (6.5, 3) circle (8pt) ;

      \node (A) at (5.5, 4.5) {(a)};
    \end{tikzpicture}
  \end{minipage}
  \hfill
  \begin{minipage}[c]{.45\linewidth}
    \begin{tikzpicture}[scale=0.4]
      \begin{scope}
        \clip (1, 0) rectangle (10, -1) ;
        \draw[thick] (1,0) -- (10,0) ;
        \foreach \x in {0, 0.25, ..., 11}
        \draw (\x, -1) -- ($ (\x, 0) + (1, 0) $) ;
      \end{scope}
      \draw[thick, decorate, decoration=snake, double] (3.5, 0) -- (3.5, 3.5) ;
      \draw[thick, decorate, decoration=snake, double] (4.5, 0) -- (4.5, 3.5) ;
      \draw[thick, decorate, decoration=snake, double] (6.5, 0) -- (6.5, 3.5) ;
      \draw[thick, decorate, decoration=snake, double] (7.5, 0) -- (7.5, 3.5) ;
      \draw[fermion,    very thick] (9, 1.5) node [anchor=west] {$b_R$}   -- (6.5, 3.5) ;
      \draw[fermionbar, very thick] (9, 5.5) node [anchor=west] {$b^\dagger_R$} -- (6.5, 3.5) ;
      \draw[fermionbar, very thick] (9, 2.5) node [anchor=west] {$t^\dagger_L$} -- (7.5, 3.5) ;
      \draw[fermion,    very thick] (9, 4.5) node [anchor=west] {$t_L$}   -- (7.5, 3.5) ;

      \draw[fermion,    very thick] (2, 1.5) node [anchor=east] {$t_R$}   -- (4.5, 3.5) ;
      \draw[fermionbar, very thick] (2, 5.5) node [anchor=east] {$t^\dagger_R$} -- (4.5, 3.5) ;
      \draw[fermionbar, very thick] (2, 2.5) node [anchor=east] {$b^\dagger_L$} -- (3.5, 3.5) ;
      \draw[fermion,    very thick] (2, 4.5) node [anchor=east] {$b_L$}   -- (3.5, 3.5) ;

      \filldraw[fill=black!50] (6.5, 3.5) circle (8pt) ;
      \filldraw[fill=black!50] (7.5, 3.5) circle (8pt) ;
      \filldraw[fill=black!50] (4.6, 3.5) circle (8pt) ;
      \filldraw[fill=black!50] (3.6, 3.5) circle (8pt) ;

      \node (B) at (5.5, 5) {(b)};
    \end{tikzpicture}
  \end{minipage}
  \caption{The gravitational interactions that give the effective
    lagrangian terms in equation (\ref{eq:quartic-lagrangian}).
    Double wiggly lines are gravitons, blobs are the radiatively
    corrected graviton-fermion vertices, and the shaded area at the
    bottom denotes the condensate. Panel (a) shows the interactions
    that, after a Fierz rearrangement give the first term of
    equation (\ref{eq:quartic-lagrangian}). Panel (b) illustrates
    the interactions that give the sixth and seventh terms of
    (\ref{eq:quartic-lagrangian}). Because the calculation is
    nonperturbative, the contribution of (b) is not necessarily
    less than that of (a).}
  \label{fig:simple-quartic}
\end{figure}
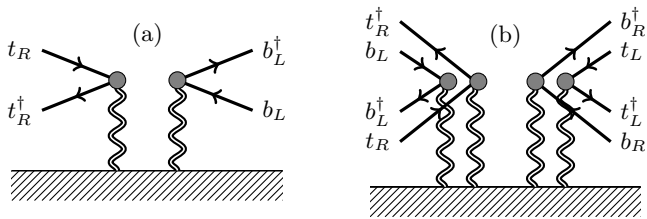

The interaction (\ref{eq:quartic-lagrangian}) implicitly assumes that
a condensate forms.  The gravitational interaction is diagonal in left
and right-handed fields, so if the interaction were perturbative, in
each term the $t$ or $b$ fields would appear with the same handedness.
The terms multiplied by $\beta$ violate this condition and require the
condensate to act as a reservoir of the appropriate helicity
states. When a condensate is present all terms can be generated, as
shown schematically in Fig.~\ref{fig:simple-quartic}.

For the quartic interaction (\ref{eq:quartic-lagrangian}) to be
diagonalizable by a Hubbard-Stratonovich shift, the constants $\alpha$, $\beta$
and $\gamma$ must obey $\gamma/\alpha = {(\beta/\alpha)}^2$.  This allows factoring the
quartic into products of quark and scalar fields whose couplings
reproduce those of the standard model.  The radiative corrections due
to weak hypercharge, to lowest order, satisfy this condition.  To see
this, consider the hypercharge factor in the correction to the
gravitational vertices of Fig.~\ref{fig:simple-quartic}(a).  This
correction gives the factor $\alpha$.  The left vertex contributes a factor
of ${(4/3)}^2$ and the right vertex ${(1/3)}^2$, their product gives
$\alpha = 16/81$.  (All radiative corrections have a common momentum factor
which is absorbed into the overall coupling, as is the interaction
with the condensate.)  Substituting the corresponding fields from the
last term of (\ref{eq:quartic-lagrangian}), the diagram
Fig.~\ref{fig:simple-quartic}(a) gives $\gamma = 4/81$.  The correction
factor $\beta$ is from the diagram in Fig.~\ref{fig:simple-quartic}(b).
The product of hypercharges in that diagram is ${(8/81)}^2$.  This
factor is shared between two of the middle terms of
(\ref{eq:quartic-lagrangian}), and I divide it equally between the
terms, giving $\beta = 8/81$.  These assignments of $\alpha$, $\beta$ and $\gamma$
satisfy the required condition.

This argument is the least satisfactory part of what is presented
here.  It is rooted in perturbation theory, and the rest of the
calculation is non-perturbative.  However, it is very suggestive and
the results below indicate that it contains some truth. I conjecture
that the full radiative corrections of weak hypercharge follow the
same pattern.  In what follows, I extract a factor of $\alpha$, absorbing
it into the overall coupling $f_t^2$, and set $\beta / \alpha = \eta$ and $\gamma / \alpha = \eta^2$, for
some positive $\eta$.  The constant $\eta$ will be the ratio of the
bottom quark mass to the top quark masses.

Diagonalizing this model starts by adding scalar fields to the
lagrangian,
\begin{equation}
  \label{eq:lagrangian-with-scalars}
  \lagrangian{} \rightarrow \frac{\eqmathbox[E]{\phiZeroDagger}
                  \eqmathbox[E]{\phiZero} +
                  \eqmathbox[E]{\phiPlus} \eqmathbox[E]{\phiMinus}}{f_t^2} +
                  \lagrangian{} \, .
\end{equation}
In principle, the new term could be multiplied by a constant $\xi > 0$
without changing any results.  I will show elsewhere that if this
model is parameterized so that it reproduces the lowest order results
of the standard model, for example, $m_W = g v / 2$, then $\xi = 1$.
Shifting the scalar fields eliminates the quartic term:
\begin{equation}
\begin{aligned}
  &\eqmathbox[A]{\phiZero} \rightarrow\, \eqmathbox[A]{\phiZero} - f_t^2 \left( \leftMass{t}{t} + \eta\, \rightMass{b}{b} \right) \, ,\\
  &\eqmathbox[A]{\phiZeroDagger} \rightarrow\, \eqmathbox[A]{\phiZeroDagger} - f_t^2 \left( \rightMass{t}{t} + \eta\, \leftMass{b}{b} \right) \, , \\
  &\eqmathbox[A]{\phiPlus} \rightarrow\, \eqmathbox[A]{\phiPlus} + f_t^2 \left( \leftMass{b}{t} - \eta\, \rightMass{b}{t} \right) \, ,\\
  &\eqmathbox[A]{\phiMinus} \rightarrow\, \eqmathbox[A]{\phiMinus} + f_t^2 \left( \rightMass{t}{b} - \eta\, \leftMass{t}{b} \right) \,  .
\end{aligned}
\end{equation}
The shifted fields are chosen so that $\Phi = {(\phiPlus, \phiZero)}^T$
transforms under $SU{(2)}_W \cross U{(1)}_Y$ with the correct quantum
numbers. Using cartesian components and splitting the Higgs field
split into the sum of the observable field $H$ and the vacuum
expectation value $v$,
\begin{align}
  \eqmathbox[C]{\phiZero} &= \frac{H + v + i w_3}{\sqrt{2}} \, ,&
  \eqmathbox[C]{\phiPlus} &= \frac{w_1 + i w_2}{\sqrt{2}} \, ,
\end{align}
the lagrangian becomes
\begin{equation}
  \begin{split}
  \label{eq:hs-cartesian-lagrangian-shifted}
    \lagrangian{} = \, & \lagrangian{fermi} +
                         \frac{H^2 + 2 H v + v^2 + w_{i} w_{i}}{2 \,  f_t^2} \, -\\
                       & \frac{v}{\sqrt{2}} \left(\rightMass{t}{t} + \leftMass{t}{t} \right) -
                         \frac{\eta v}{\sqrt{2}} \left(\rightMass{b}{b} + \leftMass{b}{b} \right) - \\
    \biggl\{ & \frac{H}{\sqrt{2}}\left[\rightMass{t}{t} + \leftMass{t}{t}
                      + \eta \left(\rightMass{b}{b} + \leftMass{b}{b} \right)\right] + \\
         i\, & \frac{w_3}{\sqrt{2}}\left[\rightMass{t}{t} - \leftMass{t}{t}
                      + \eta \left(\leftMass{b}{b} - \rightMass{b}{b} \right)\right] - \\
             & \frac{w_1}{\sqrt{2}}\left[\rightMass{t}{b} + \leftMass{b}{t}
                      - \eta \left(\leftMass{t}{b} + \rightMass{b}{t} \right)\right] - \\
         i\, & \frac{w_2}{\sqrt{2}}\left[\rightMass{t}{b} - \leftMass{b}{t}
                      - \eta \left(\leftMass{t}{b} - \rightMass{b}{t}\right)\right] \biggr\} \, .
  \end{split}
\end{equation}
It is apparent that at the semiclassical level the top and bottom
quarks acquire masses of $v / \sqrt{2}$ and $\eta v / \sqrt{2}$, respectively,
justifying the statement that $\eta$ is ratio of the quark masses, and the
interpretation of the $t$ and $b$ fields as the top and bottom quarks.

As yet, the scalar excitations of the condensate have no kinetic
terms.  Well below the transition temperature, I can integrate out the
fermions to get an effective action for the scalars alone, which will
have the expected kinetic terms.

Collecting the fermions into a Nambu spinor,
\begin{equation}
  \Psi(p) = {\left[ t_L(p), t_R(p), b_L(p), b_R(p) \right]}^T \, ,
\end{equation}
the action is
\begin{equation}
  S = \int \shortMomentumMeasure{p} \, \shortMomentumMeasure{p'}\,
  \Psi^\dagger(p) M(p, p') \Psi(p') \, ,
\end{equation}
where
\begin{equation}
  \begin{aligned}
    M(p, p') & = G^{-1}(p) + u(p, p') \, ,\\
    G^{-1}(p) & = \diag \left[ \sigmabar \cdot p, \sigma \cdot p,
                \sigmabar \cdot p, \sigma \cdot p \right] \, ,
  \end{aligned}
\end{equation}
and
\begin{equation}
  \begin{aligned}
    & u  = \\
    & \begin{bmatrix}
       0  & -\frac{H + v + i w_3}{\sqrt{2}} & 0  & \frac{\eta (w_1 - i w_2)}{\sqrt{2}} \\
       -\frac{H + v - i w_3}{\sqrt{2}} & 0 & -\frac{w_1 - i w_2}{\sqrt{2}} & 0 \\
       0 & -\frac{w_1 + i w_2}{\sqrt{2}} & 0  & -\frac{\eta (H + v - i w_3)}{\sqrt{2}} \\
       \frac{\eta (w_1 + i w_2)}{\sqrt{2}} & 0 & -\frac{\eta (H + v + i w_3)}{\sqrt{2}} & 0
    \end{bmatrix} \, .
  \end{aligned}
\end{equation}
Integrating out the fermi field $\Psi$ gives an action for the scalars:
\begin{equation}
  \begin{aligned}
    S_{\mathrm{eff}} = \int \shortMomentumMeasure{p} \,
    & \frac{1}{2 f_t^2} (H^2 + 2Hv + v^2 + w_i w_i) \, - \\
    & i \Tr \log (\mathbb{1} + Gu) \, ,
    \label{eq:integrated-effective-action}
  \end{aligned}
\end{equation}
where a factor of $G^{-1}{(p)}$ has been used to make the argument of
the logarithm dimensionless at the expense of a change in the
normalization of the functional integral.  The `$\Tr$' symbol denotes
the combined operator and matrix trace, `$\tr$' will be the matrix
trace alone.

It is possible to treat $Gu$ in equation
(\ref{eq:integrated-effective-action}) as small and expand the
logarithm. However, this would be an expansion about the symmetric
vacuum, valid only for energies near the electroweak transition.  To
find the low energy effective action (and therefore the observable
particle spectrum), an expansion about the true (condensed) vacuum is
needed. The first step is to find the condensate density by
extremizing the action with respect to $v$:
\begin{equation}
  \label{eq:gap-equation}
  \begin{aligned}
    0 & = \fderiv{S_{\mathrm{eff}}}{v} \\
    & = \int \shortMomentumMeasure{p} \left\{ \frac{v}{ f_t^2} - i \, N_c \bra{p} \tr  \left[ \Delta \left.\fderiv{u}{v}\right|_0 \right] \ket{p} \right\} \, ,
  \end{aligned}
\end{equation}
where $\Delta = {(\mathbb{1} + G u_0)}^{-1} G$, $N_c$ is the number of
quark colors and a zero subscript indicates evaluation at $H = w_i =
0$.  Inserting a resolution of the identity as free particle momentum
states between $\Delta$ and $\fderivText{u}{v}$ gives
\begin{equation} 0 = \frac{1}{2 N_c f_t^2} + i \int
\shortMomentumMeasure{p} \, \left[ \frac{1}{p^2 - \frac{v^2}{2}} +
\frac{\eta^2}{p^2 - \frac{\eta^2 v^2}{2}} \right] \, .
\end{equation}
This is the equation for $v$, analogous to the gap equation of
superconductivity.  The integral is divergent, but it is not necessary
to evaluate it. Instead, in what follows, it is used to eliminate
$f_t$, which is not accessible to measurement, in favor of $v$, which
is.

The kinetic term for the Higgs field is the piece quadratic in $H$ in
the expansion of the effective action,
\begin{equation}
  \int \shortMomentumMeasure{r} \shortMomentumMeasure{s} \,
  \frac{1}{2} H(r) \fderivTwo{S_{\mathrm{eff}}}{H(r)}{H(s)} H(s) \, ,
\end{equation}
which, in the parameterization of the standard model, is
\begin{equation}
  \label{eq:standard-model-higgs-kinetic}
  \int \shortMomentumMeasure{r} \, \frac{Z(r^2)}{2} H(-r) (r^2 - M^2_{H}) H(r) \, .
\end{equation}
Because the kinetic term is generated by quantum effects, a field
renormalization factor, $Z$, must be anticipated.

The functional derivative is
\begin{equation}
  \label{eq:effective-higgs-kinetic}
  \begin{aligned}
    & \fderivTwo{S_{\mathrm{eff}}}{H(r)}{H(s)} = \\
    & \frac{1}{f_t^2} + i \int \shortMomentumMeasure{p} \, \bra{p}
      \tr \left[ \Delta(p) \fderiv{u}{H(r)} \Delta(p - r) \fderiv{u}{H(s)} \right] \ket{p} \, .
  \end{aligned}
\end{equation}
The first term on the right hand side can be eliminated by using the
gap equation and the integral reduced to Feynman integrals as the gap
equation was, giving
\begin{equation}
   \label{eq:root-integral}
   \begin{aligned}
     & \fderivTwo{S_{\mathrm{eff}}}{H(r)}{H(s)} = & \\
     & - i \, \shortDelta{r + s} \int_0^1 dx \int \shortMomentumMeasure{p}
     & \frac{r^2 - 2 v^2}{{\left[ p^2 - \frac{\vphantom{\eta^2} v^2}{2} + r^2 x (1 - x) \right]}^2} \, + \\
     & & \frac{\eta^2 (r^2 - 2 \eta^2 v^2)}{{\left[ p^2 - \frac{\eta^2 v^2}{2} + r^2 x (1 - x) \right]}^2} \, .
   \end{aligned}
 \end{equation}
If $\eta = 0$, the bottom quark decouples and the above has a real root
$r = \pm\sqrt{2}v$, twice the mass of the top quark.  If $\eta \ne 0$, it has
no roots on the physical sheet, but there are complex roots on other
Riemann sheets.  The roots come in groups of four, with equal magnitude:
positive and negative energy, and growing and decaying modes (these
quartets of roots were observed for the collective modes of BCS by
Andrianov and Popov~\cite{andrianov:1976}).  The complex roots are on
unphysical sheets because unitarity requires that the action be
real~\cite{eden:1966}.

The integral in (\ref{eq:root-integral}) is regularized using a
Pauli-Villars parameter $\Lambda$.  Introducing dimensionless parameters
$\rho^2 = r^2 / v^2$ and $m_H^2 = M_H^2 / v^2$, and equating the standard
model Higgs kinetic term (\ref{eq:standard-model-higgs-kinetic}) to
that of the effective model (\ref{eq:effective-higgs-kinetic}), gives
\begin{equation}
  \label{eq:evaluated-integral}
  \begin{aligned}
  Z(\rho^2)  (\rho^2 - m_H^2) = \, (\rho^2 - 2) & I_n(\rho, 1) \, + \\
    \eta^2 (\rho^2 - 2 \eta^2) & I_n(\rho, \eta) \, + \\
    & C_{\Lambda} \, ,
  \end{aligned}
\end{equation}
where $C_{\Lambda}$ is the regularized divergent term and the finite
part of the integral is
\begin{equation}
  \begin{aligned}
    \label{eq:root-function}
    I_n(\rho, & \eta) = \log \left(\eta^2 / 2 \right) - 2 \, + \\
           & \frac{2}{\rho} \sqrt{2 \eta^2 - \rho^2} \left[ \arctan \left(\frac{\rho}{\sqrt{2 \eta^2 - \rho^2}} \right) + n \pi \right] \, .
  \end{aligned}
\end{equation}
In this expression, $n$ is the sheet index.  Using on-shell
renormalization, the equation whose root determines the Higgs mass is
\begin{equation}
  \begin{aligned}
    \label{eq:root}
    0 = \, & (\rho^2 - 2) I_n(\rho, 1) + \eta^2 (\rho^2 - 2 \eta^2) I_n(\rho, \eta) \, - \\
       \Re \bigl[ & (m_H^2 - 2) I_1(m_H, 1) + \eta^2 (m_H^2 - 2 \eta^2) I_1(m_H, \eta) \bigr] \, .
  \end{aligned}
\end{equation}
The second line of the above is the counterterm that fixes the real
part of the excitation on the first unphysical sheet to be the mass of
the observed Higgs boson.  (The imaginary part can not be specified
since it is determined by unitarity.)  Note that this expression is
independent of the number of quark colors, $N_c$.  The Higgs mass does
depend on the number of colors, but this dependence is contained
entirely in the vacuum expectation value $v$.

On the right hand side of equation (\ref{eq:root}), before the
counterterm, are a term multiplied by $\eta = 1$ (the ``top term'') and
another multiplied by $\eta \ne 1$ (the ``bottom term''). Roots exist if
only top term is off the physical sheet ($n \ge 1$) or if both the top
and bottom terms are off the physical sheet.  The roots are close to
the real-axis branch cuts for both the bottom and top terms, and the
numerical results given below put both top and bottom terms on the
same sheet.  However, roots still exist if only the top term is on a
non-physical sheet and the bottom term remains on the physical ($n =
0$) sheet.  In that case, the real part of the root is only slightly
different from when both terms are on the same sheet, but the
magnitude of the imaginary part is significantly smaller.

For numerical evaluation, I take the ratio of the bottom to top mass
to be $\eta = 0.015$, corresponding to the ratio of the running bottom
mass measured at the energy scale of the Higgs boson,
$m_b(m_H)$~\cite{aparisi:2022} and the measured top quark pole
mass~\cite{navas:2024}.  The typical residual of the numerical root
finder is a few parts in $10^{15}$.

\begin{figure}
  %
%
%
%
\begin{tikzpicture}[gnuplot]
\path (0.000,0.000) rectangle (8.572,5.080);
\gpcolor{color=gp lt color border}
\gpsetlinetype{gp lt border}
\gpsetdashtype{gp dt solid}
\gpsetlinewidth{2.00}
\draw[gp path] (1.504,0.985)--(1.684,0.985);
\draw[gp path] (8.019,0.985)--(7.839,0.985);
\node[gp node right] at (1.320,0.985) {$0.01$};
\draw[gp path] (1.504,1.270)--(1.594,1.270);
\draw[gp path] (8.019,1.270)--(7.929,1.270);
\draw[gp path] (1.504,1.437)--(1.594,1.437);
\draw[gp path] (8.019,1.437)--(7.929,1.437);
\draw[gp path] (1.504,1.555)--(1.594,1.555);
\draw[gp path] (8.019,1.555)--(7.929,1.555);
\draw[gp path] (1.504,1.647)--(1.594,1.647);
\draw[gp path] (8.019,1.647)--(7.929,1.647);
\draw[gp path] (1.504,1.722)--(1.594,1.722);
\draw[gp path] (8.019,1.722)--(7.929,1.722);
\draw[gp path] (1.504,1.785)--(1.594,1.785);
\draw[gp path] (8.019,1.785)--(7.929,1.785);
\draw[gp path] (1.504,1.840)--(1.594,1.840);
\draw[gp path] (8.019,1.840)--(7.929,1.840);
\draw[gp path] (1.504,1.888)--(1.594,1.888);
\draw[gp path] (8.019,1.888)--(7.929,1.888);
\draw[gp path] (1.504,1.932)--(1.684,1.932);
\draw[gp path] (8.019,1.932)--(7.839,1.932);
\node[gp node right] at (1.320,1.932) {$0.1$};
\draw[gp path] (1.504,2.216)--(1.594,2.216);
\draw[gp path] (8.019,2.216)--(7.929,2.216);
\draw[gp path] (1.504,2.383)--(1.594,2.383);
\draw[gp path] (8.019,2.383)--(7.929,2.383);
\draw[gp path] (1.504,2.501)--(1.594,2.501);
\draw[gp path] (8.019,2.501)--(7.929,2.501);
\draw[gp path] (1.504,2.593)--(1.594,2.593);
\draw[gp path] (8.019,2.593)--(7.929,2.593);
\draw[gp path] (1.504,2.668)--(1.594,2.668);
\draw[gp path] (8.019,2.668)--(7.929,2.668);
\draw[gp path] (1.504,2.731)--(1.594,2.731);
\draw[gp path] (8.019,2.731)--(7.929,2.731);
\draw[gp path] (1.504,2.786)--(1.594,2.786);
\draw[gp path] (8.019,2.786)--(7.929,2.786);
\draw[gp path] (1.504,2.835)--(1.594,2.835);
\draw[gp path] (8.019,2.835)--(7.929,2.835);
\draw[gp path] (1.504,2.878)--(1.684,2.878);
\draw[gp path] (8.019,2.878)--(7.839,2.878);
\node[gp node right] at (1.320,2.878) {$1$};
\draw[gp path] (1.504,3.163)--(1.594,3.163);
\draw[gp path] (8.019,3.163)--(7.929,3.163);
\draw[gp path] (1.504,3.330)--(1.594,3.330);
\draw[gp path] (8.019,3.330)--(7.929,3.330);
\draw[gp path] (1.504,3.448)--(1.594,3.448);
\draw[gp path] (8.019,3.448)--(7.929,3.448);
\draw[gp path] (1.504,3.540)--(1.594,3.540);
\draw[gp path] (8.019,3.540)--(7.929,3.540);
\draw[gp path] (1.504,3.615)--(1.594,3.615);
\draw[gp path] (8.019,3.615)--(7.929,3.615);
\draw[gp path] (1.504,3.678)--(1.594,3.678);
\draw[gp path] (8.019,3.678)--(7.929,3.678);
\draw[gp path] (1.504,3.733)--(1.594,3.733);
\draw[gp path] (8.019,3.733)--(7.929,3.733);
\draw[gp path] (1.504,3.781)--(1.594,3.781);
\draw[gp path] (8.019,3.781)--(7.929,3.781);
\draw[gp path] (1.504,3.825)--(1.684,3.825);
\draw[gp path] (8.019,3.825)--(7.839,3.825);
\node[gp node right] at (1.320,3.825) {$10$};
\draw[gp path] (1.504,4.109)--(1.594,4.109);
\draw[gp path] (8.019,4.109)--(7.929,4.109);
\draw[gp path] (1.504,4.276)--(1.594,4.276);
\draw[gp path] (8.019,4.276)--(7.929,4.276);
\draw[gp path] (1.504,4.394)--(1.594,4.394);
\draw[gp path] (8.019,4.394)--(7.929,4.394);
\draw[gp path] (1.504,4.486)--(1.594,4.486);
\draw[gp path] (8.019,4.486)--(7.929,4.486);
\draw[gp path] (1.504,4.561)--(1.594,4.561);
\draw[gp path] (8.019,4.561)--(7.929,4.561);
\draw[gp path] (1.504,4.624)--(1.594,4.624);
\draw[gp path] (8.019,4.624)--(7.929,4.624);
\draw[gp path] (1.504,4.679)--(1.594,4.679);
\draw[gp path] (8.019,4.679)--(7.929,4.679);
\draw[gp path] (1.504,4.728)--(1.594,4.728);
\draw[gp path] (8.019,4.728)--(7.929,4.728);
\draw[gp path] (1.504,4.771)--(1.684,4.771);
\draw[gp path] (8.019,4.771)--(7.839,4.771);
\node[gp node right] at (1.320,4.771) {$100$};
\draw[gp path] (1.504,0.985)--(1.504,1.165);
\draw[gp path] (1.504,4.771)--(1.504,4.591);
\node[gp node center] at (1.504,0.677) {0};
\draw[gp path] (3.133,0.985)--(3.133,1.165);
\draw[gp path] (3.133,4.771)--(3.133,4.591);
\node[gp node center] at (3.133,0.677) {0.01};
\draw[gp path] (4.762,0.985)--(4.762,1.165);
\draw[gp path] (4.762,4.771)--(4.762,4.591);
\node[gp node center] at (4.762,0.677) {0.02};
\draw[gp path] (6.390,0.985)--(6.390,1.165);
\draw[gp path] (6.390,4.771)--(6.390,4.591);
\node[gp node center] at (6.390,0.677) {0.03};
\draw[gp path] (8.019,0.985)--(8.019,1.165);
\draw[gp path] (8.019,4.771)--(8.019,4.591);
\node[gp node center] at (8.019,0.677) {0.04};
\draw[gp path] (1.504,4.771)--(1.504,0.985)--(8.019,0.985)--(8.019,4.771)--cycle;
\gpsetdashtype{gp dt 2}
\draw[gp path](1.505,3.197)--(8.020,3.197);
\node[gp node center,rotate=-270] at (0.292,2.878) {Decay width to $b \bar{b}$ (MeV)};
\node[gp node center] at (4.761,0.215) {$\eta = m_b / m_t$};
\gpsetdashtype{gp dt solid}
\draw[gp path] (1.667,1.149)--(1.830,1.719)--(1.993,2.052)--(2.156,2.289)--(2.318,2.472)%
  --(2.481,2.622)--(2.644,2.749)--(2.807,2.859)--(2.970,2.956)--(3.133,3.042)--(3.296,3.120)%
  --(3.459,3.192)--(3.621,3.258)--(3.784,3.318)--(3.947,3.375)--(4.110,3.428)--(4.273,3.478)%
  --(4.436,3.525)--(4.599,3.569)--(4.762,3.611)--(4.924,3.651)--(5.087,3.689)--(5.250,3.726)%
  --(5.413,3.761)--(5.576,3.794)--(5.739,3.826)--(5.902,3.857)--(6.065,3.887)--(6.227,3.916)%
  --(6.390,3.943)--(6.553,3.970)--(6.716,3.996)--(6.879,4.021)--(7.042,4.046)--(7.205,4.069)%
  --(7.368,4.092)--(7.530,4.115)--(7.693,4.136)--(7.856,4.158)--(8.019,4.178);
\draw[gp path] (1.504,4.771)--(1.504,0.985)--(8.019,0.985)--(8.019,4.771)--cycle;
\gpdefrectangularnode{gp plot 1}{\pgfpoint{1.504cm}{0.985cm}}{\pgfpoint{8.019cm}{4.771cm}}
\end{tikzpicture}
  \caption{Higgs partial width to $b \bar{b}$ as a function
    of the effective mass ratio $\eta = m_t/ m_b$.  The dashed
    horizontal line is the partial width of the standard model Higgs
    to $b \bar{b}$ taken from~\cite{almeida:2014}.}
  \label{fig:etaSweep}
\end{figure}
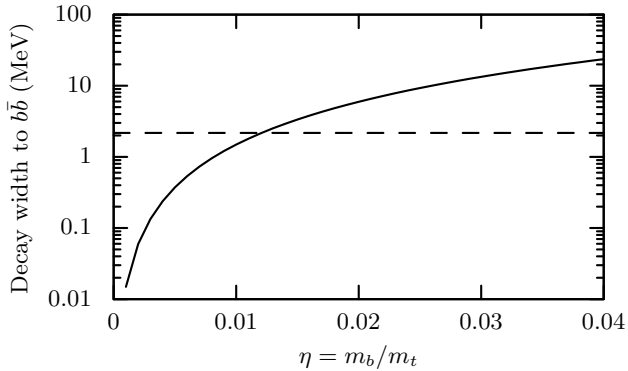
The partial width to $b \bar{b}$ for $\eta = 0.015$ is 3.35 MeV, about
fifty percent larger the standard model prediction for a pointlike
Higgs~\cite{atlas_width:2023, cms_width:2022}.  It is not surprising
that the Higgs decay rate here is different than in the standard
model, since here the Higgs is not a fundamental scalar particle.  Its
mass is about half of the condensate density $v$, an energy scale at
which the constituents of the condensate may begin to be resolved. For
other values of $\eta$ the partial width to $b \bar{b}$ is shown in
Fig.~(\ref{fig:etaSweep}).
\begin{figure}[b]
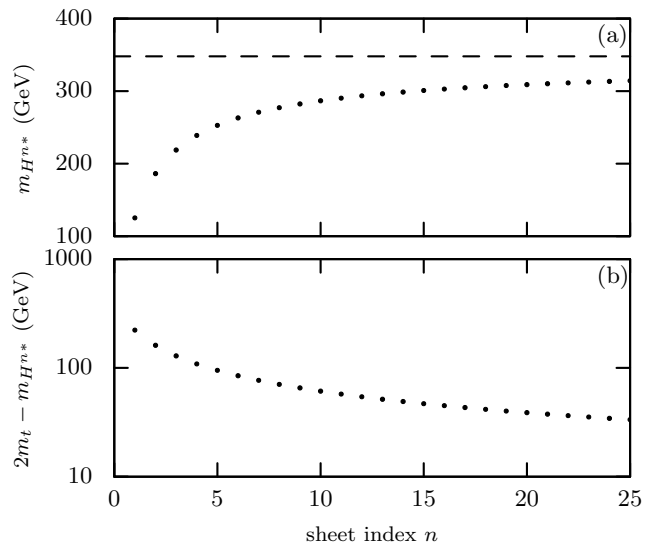

  \include{sheetSweep.tex}
  \caption{Panel (a): Masses of the first twenty five Higgs excitations
    vs.\ sheet index.  The point corresponding to $n = 1$ is the 125 GeV
    Higgs boson. The dashed horizontal line is at 347.88 GeV, the root of the
    Higgs mass equation for $n = 10^6$, $\eta = 0.015$.
    Panel (b): The ``binding energy'', i.e., difference in mass between a Higgs
    excitation and the asymptotic mass of the Higgs excitations, where the
    last has been approximated by the mass on the sheet with $n = 10^6$.
    Over the range  $10 \le n \le 10^4$, the binding energy is approximately
    proportional to $n^{-2/3}$.}
  \label{fig:sheetSweep}
\end{figure}
The first few roots of (\ref{eq:root}) for $n > 1$ are given in
Table~\ref{tbl:excitations}, along with their estimated decay widths,
assuming a pointlike Higgs with standard model couplings to $W^+W^-$
and $ZZ$.  Numerical examination of the roots of (\ref{eq:root})
indicates that they are all simple, corresponding to isolated poles in
the propagator.

The real parts of the first twenty five roots are plotted in
Fig.~(\ref{fig:sheetSweep}). The top panel plots the masses and the
bottom panel shows the ``binding energy'', the difference between the
mass and the asymptotic mass for large sheet index ($ \approx 2 m_t$).
\begin{table}[t]
  \begin{ruledtabular}
    \caption{The masses and estimated partial widths to $W^{+}W^{-}$ and
      $ZZ$ of the first six Higgs excitations.  For excitations
      beyond $H^{5*}$, the width is comparable to or larger than the spacing
      between excitations, so they are not distinct.
      The third column is the partial width to weak vector bosons
      assuming a pointlike Higgs boson and standard model $HZZ$
      and $HWW$ coupling strengths.\label{tbl:excitations}
    }%
    \begin{tabular}{ccc}
      Excitation & Mass (GeV) & Width (GeV) \\
      \hline
      $H^{*}$  & 186.34 & 0.88 \\
      $H^{**}$ & 218.82 & 2.22 \\
      $H^{3*}$ & 239.00 & 3.30 \\
      $H^{4*}$ & 252.89 & 4.20 \\
      $H^{5*}$ & 263.10 & 4.95 \\
      $H^{6*}$ & 270.97 & 5.59 \\
    \end{tabular}
  \end{ruledtabular}
\end{table}
That the binding energy of the Higgs excitations decreases as
$n^{-2/3}$ is evidence against interpreting the Higgs as a kind of
bound state of valence quarks.  This model has no gauge fields and
the gravitational interaction is treated in the random phase
approximation~\cite{anderson:1958}, so there is no potential between
fermionic field oscillators.  A physical interpretation of the Higgs
excitations might be itinerant ``bubbles'' or dislocations in the
ordered ground state.

What I have done above parallels the work of Andrianov and
Popov~\cite{andrianov:1976} and Popov~\cite{popov:1987} in which they
computed the collective excitations of the condensate in BCS
superconductivity.  They predicted a series of Higgs-like amplitude
excitations in superconductors which are unfortunately hard to detect
because there is no external probe that couples linearly to the Higgs
modes of a superconductor~\cite{shimano:2020}.  In contrast, the Higgs
excitations described here appear to have the same coupling as the
standard model Higgs, and are at energies in reach of existing
accelerators, allowing their existence to be checked directly.


\begin{acknowledgments}
  I would like to thank R.W. Wilson for listening to these ideas at an
  early stage and the late R.W. Lucky for additional discussions.
  At Nokia Bell Labs, Wolfgang Templ suggested the problem that led
  to this work, and Timo Soirinsuo provided continued support.  Their help is
  gratefully acknowledged.
\end{acknowledgments}


\bibliography{higgs-excitations.bib}

\end{document}